\documentstyle[amsfonts,epsf,11pt]{article}

\def\baselinestretch{1.2}
\newcommand{\be}{\begin{equation}}
\newcommand{\ee}{\end{equation}}

\begin{document}

\title{{\bf Fractal and complex network analyses of protein molecular dynamics}}
\author{Yuan-Wu Zhou$^{1,2}$, Jin-Long Liu$^1$, Zu-Guo Yu$^{1,3}$\thanks{Corresponding to yuzuguo@aliyun.com},\ \  Zhi-Qin Zhao$^1$  and Vo Anh$^3$\\
{\small $^1$Hunan Key Laboratory for Computation and Simulation in
Science and Engineering and }\\
{\small Key Laboratory of Intelligent Computing and Information
Processing of Ministry of Education,}\\
{\small  Xiangtan University,
Xiangtan, Hunan 411105, China.}\\
{\small $^2$Civil Construction Engineering Department, Guangxi
University of Science and Technology,}\\
{\small Liuzhou 545000, China.}\\
{\small $^3$School of Mathematical Sciences, Queensland University
of Technology, GPO Box 2434,} \\
{\small Brisbane, Q4001, Australia.}}
\date{}
\maketitle

\begin{abstract}
Based on protein molecular dynamics, we investigate the fractal
properties of energy, pressure and volume time series using the
multifractal detrended fluctuations analysis (MF-DFA) and the
topological and fractal properties of their converted horizontal
visibility graphs (HVGs). The energy parameters of protein
dynamics we considered are bonded potential, angle potential,
dihedral potential, improper potential, kinetic energy, Van der
Waals potential, electrostatic potential, total energy and
potential energy. The shape of the $h(q)$ curves from MF-DFA
indicates that these time series are multifractal. The numerical
values of the exponent $h(2)$ of MF-DFA show that the series of
total energy and potential energy are non-stationary and
anti-persistent; the other time series are stationary and
persistent apart from series of pressure (with $H\approx 0.5$
indicating the absence of long-range correlation). The degree
distribution of their converted HVGs show that these networks are
exponential. The results of fractal analysis show that fractality
exists in these converted HVGs. For each energy, pressure or
volume parameter, it is found that the values of $h(2)$ of MF-DFA
on the time series, exponent $\lambda$ of the exponential degree
distribution and fractal dimension $d_B$ of their converted HVGs
do not change much for different proteins (indicating some
universality). We also found that after taking average over all
proteins, there is a linear relationship between $\langle h(2)
\rangle$ (from MF-DFA on time series) and $\langle d_{B} \rangle$
of the converted HVGs for different energy, pressure and volume.
\end{abstract}


\section{Introduction}

\ \ \ \  Proteins are among the most important biomacromolecules
because they can perform biological functions, which are usually
determined by their structures. To date, structures of 97362
proteins (updated in Protein Data Bank \cite{Berman00} on
2014-01-28) have been obtained by experimental methods, such as
X-ray $(88.4\%)$, solution NMR $(10.6\%)$ and electron microscopy
$(0.7\%)$.  On the other hand, molecular dynamics simulation
packages such as NAMD ({\it Not (just) Another Molecular Dynamics
program}) \cite{Phillips05} has been developed to learn about
different aspects of proteins. Protein molecule energy includes
kinetic energy and potential energy, and potential energy can be
calculated by quasi-empirical way (such as CHARMM \cite{Brooks09}
and AMBER \cite{Case05}). In fact, Rueda {\it et al.}
\cite{Rueda07} used NAMD to analyze the molecular dynamics of 30
real proteins. This paper uses NAMD to simulate protein molecular
dynamics, then derive the time series for some corresponding
parameters whose features reflect the aspects of protein
structures.

Apart from being characterized by an enormous number of degrees of
freedom, proteins have multidimensional potential energy surfaces
\cite{Banerji11}. It is well known that the self-similarity
exhibiting in the distributions of their biophysical and
biochemical properties can serve as an effective tool to extract
the inherently inhomogeneous and nonlinear behaviors of protein
structures. Fractal dimension (FD) has been widely used to
characterize the self-similarity of fractal objects
\cite{Mandelbrot1983,Falconer1997}. Many FD-based methods have
been proposed to investigate protein structures \cite{Banerji11}.
Fractal methods can also be used to characterize the scaling
properties of time series and then to reveal the self-similarity
of the original system \cite{Mandelbrot1983}. A multifractal
system is a generalization of a fractal system in which a single
exponent (the fractal dimension) is not enough to describe its
dynamics; instead, a continuous spectrum of exponents (the
so-called singularity spectrum) is needed \cite{Harte2001}. The
concept of multifractal phenomena describes that different regions
of an object have different fractal properties, and multifractal
scaling provides a quantitative description of a broad range of
heterogeneous phenomena \cite{Stanley1988}. Multifractal analysis
was initially proposed to treat turbulence data and is a useful
way to characterize the spatial heterogeneity of both theoretical
and experimental fractal patterns \cite{Halsey86}. The
multifractal detrended fluctuations analysis (MF-DFA)
\cite{Kantelhardt02} is an extension of the standard detrended
fluctuation analysis (DFA) introduced by Peng {\it et al}.
\cite{Peng92,Peng94}. DFA can be employed to detect long-range
correlations in stationary and nonstationary time series. Hence
MF-DFA is a suitable tool to characterize the multifractal
property and long-range correlation in time series. Multifractal
analysis has been used to study genomes [15-17], protein
structures  and functions (e.g. [18-22]). In this paper, we
investigate by MF-DFA \cite{Kantelhardt02} the scaling property of
the associated time series of energy, pressure and volume derived
from simulations of protein molecular dynamics.

In last decade, the study of complex networks has gained
prominence across many disciplines of science. Studies have shown
that complex network theory has become a powerful tool to analyze
protein complex structures [23-25]. Single proteins in 3D space
can also be considered as biological complex systems emerged
through the interactions of their constituent amino acids. These
interactions among the amino acids within a protein can be
presented as residues interaction network (RIN) (also called
residues interaction graphs (RIGs), protein contact network (PCN),
protein structure network (PSN), protein contact map (PCM), amino
acid network (AAN)), which can be constructed with varying
definitions of nodes and edges [23-25].

Recent studies showed that complex network theory (such as
recurrence networks) may also be an effective method to analyze
time series [26-33]. Lacasa {\it et al.} \cite{Lacasa08} proposed
the visibility graph (VG) algorithm to convert time series into
complex networks. Then Luque {\it et al.} \cite{Luque09} proposed
the horizontal visibility algorithm to convert time series into
complex networks. It has been shown that these converted networks
inherit several properties of time series in the structure of
networks [34-36]. Therefore, we can understand time series from a
new point of view using the converted networks. In this paper, we
hope to reveal some meaningful information in the associated time
series of energy, pressure and volume for real proteins from the
perspective of the horizontal visibility graphs (HVGs)
\cite{Luque09}. This prompts us to further study the fundamental
topological and fractal properties of the converted HVGs from
different energy, pressure and volume time series of real
proteins.

The fractal and self-similarity properties of complex network have
also been focused and studied widely in a variety of fields
[37-39]. It is found that many complex networks, including the
world-wide web (WWW), social networks, protein-protein interaction
(PPI) networks and cellular networks, are self-similar under a
certain length-scale. Some numerical algorithms have been proposed
to calculate the fractal dimensions of complex networks. Song {\it
et al.}~\cite{Song2007} proposed a box-counting algorithm to
calculate their fractal dimension. Kim {\it et al.} \cite{Kim2007}
studied the skeleton and fractal scaling in complex networks via
an improvement method. Later on, Zhou {\it et al.} \cite{Zhou07}
introduced an alternative algorithm to detect self-similarity of
cellular networks. Recently, Li {\it et al.} \cite{Li14} studied
the fractal properties of a family of fractal networks using the
random sequential box-covering algorithm proposed by Kim {\it et
al.} \cite{Kim2007}. Liu {\it et al.} \cite{Liu14} adopted this
algorithm to calculate the fractal dimensions of the recurrence
networks constructed from fractional Brownian motions. In this
paper, we also adopting the random sequential box-covering
algorithm to calculate the fractal dimension of the converted HVGs
of the time series of energy, pressure and volume for real
proteins.

\section{ Molecular dynamics for protein}

\ \ \ \  The behavior of biomolecular systems can be modelled by
molecular dynamics, which is helpful to understand protein
functions. A common procedure of molecular dynamics \cite{Kumar11}
for proteins is as follows:

 (a) Initialize the positions and velocities of all atoms. The initial positions of atoms come from Protein Data Bank (PDB \cite{Berman00}).
 Add water and ions to a box with protein, minimize the box system energy. Then, the initial velocities of atoms are
 assigned from Maxwell's distribution.

 (b) Calculate forces for all atoms using potential energy. The protein potential energy can be defined as
\begin{eqnarray}
V=& &V_{bonds}+V_{angles}+V_{dihedrals}+V_{impropers} \nonumber\\
& &+V_{Vdw}+V_{electrostatic}, \label{1}
\end{eqnarray}
where $V$ is protein potential energy, $V_{bonds}$ is bonded
potential, $V_{angles}$ is angle potential,
 $V_{dihedrals}$ is dihedral potential, $V_{impropers}$ is improper potential, $V_{Vdw}$ is Van der Waals potential,
 and $V_{electrostatic}$ is electrostatic potential. In this paper, we analyze the potential energy and kinetic energy time series.
 Then, forces that act on an atom can be written as
\begin{eqnarray}
\overrightarrow{F}=\frac{\partial V(\vec{r})}{\partial \vec{r}},
\label{2}
\end{eqnarray}
where $\vec{r}$ is the position vector of atoms.

(c) Apply thermostat and volume changes. Update positions and
velocities. According to Newton's law, the kinetic equation can be
written as
\begin{eqnarray}
\overrightarrow{F}=m\frac{\partial^{2}\vec{r}}{\partial t^{2}}.
\label{3}
\end{eqnarray}
Using an algorithm, such as the Verlet algorithm or LeapFrog
algorithm, new positions and velocities can be calculated.

(d) Repeat steps (b) and (c) till a termination condition is met.

(e).  Analyze data. NAMD is a software for biomolecular molecular
dynamics \cite{Phillips05}. Some researchers used  NAMD to achieve
some important results \cite{Rueda07}. In the paper, we use it to
simulate protein dynamics.

\section{Multifractal detrended fluctuation analysis for time series}

\ \ \ \  Detrended fluctuations analysis (DFA) introduced by Peng
{\it et al}. \cite{Peng92,Peng94} can be employed to detect
long-range correlations in stationary and nonstationary time
series. The multifractal detrended fluctuations analysis (MF-DFA)
\cite{Kantelhardt02} is an extension of DFA. This approach can be
described as follows. Given a time series $x_{k}$, $k=1, \cdots,
N$, where $N$ is the length of the series, the profile $Y(i)$ is
defined as
\begin{eqnarray}
 Y(i)\equiv \sum_{k=1}^{i}[x_{k}-\langle x\rangle]. \ \ \ i=1, \cdots, N,
\end{eqnarray}
where $\langle x\rangle$ is the mean of sequence $x_{k}$. Then
$Y(i)$ is divided into $N_{s}\equiv int(N/s)$ non-overlapping and
continuous segments of equal length $s$. Then we estimate the
variance of $y(i)$ by
\begin{eqnarray}
F^{2}(s,v)\equiv
\frac{1}{s}\sum_{i=1}^{s}\{Y[(v-1)s+i]-y_{v}(i)\}^{2}, \label{1}
\end{eqnarray}
for each segment $v$, $v=1, \cdots, N$ and
\begin{eqnarray}
F^{2}(s,v)\equiv
\frac{1}{s}\sum_{i=1}^{s}\{Y[N-(v-N_{s})s+i]-y_{v}(i)\}^{2},
\label{2}
\end{eqnarray}
for $v=N_{s}+1, \cdots, 2N_{s}$, where $y_{v}(i)$ is a fitting
polynomial in segment $v$ (linear, quadratic, cubic, or higher
order polynomial). Last we take the average over $2N_{s}$ segments
to achieve the $qth$ order fluctuation function
\begin{eqnarray}
F_{q}(s)\equiv \{\frac{1}{2N_{s}} \sum_{v=1}^{2N_{s}}
[F^{2}(s,v)]^{q/2} \}^{1/q}, \ \ \ (if \ q\neq 0). \label{3}
\end{eqnarray}
\begin{eqnarray}
F_{q}(s)\equiv exp\{\frac{1}{4N_{s}} \sum_{v=1}^{2N_{s}}
\ln[F^{2}(s,v)]\}, \ \ \ (if \ q=0). \label{4}
\end{eqnarray}
 We will assume
that $F_{q}(s)$ is characterized by a power law:%
\begin{equation}
F_{q}(s)\ \propto \ s^{h\left( q\right) }.
\end{equation}%
The scaling function $h\left( q\right) $ is then determined by the
regression of $\ln F_{q}(s)$ on $\ln s$ in some range of time
scale $s$.

MF-DFA is suitable for both stationary and nonstationary time
series \cite{Kantelhardt02}. We denote by $H$ the Hurst exponent
of time series. The range $0.5 <H < 1$ indicates persistence; and
the range $0 < H < 0.5$ indicates anti-persistence; for
uncorrelated series, the scaling exponent $H$ is equal to 0.5
\cite{Mandelbrot68}. Assuming the setting of fractional Brownian
motion, Movahed {\it et al}. \cite{Movahed06} proved the relation
$H =h(2)-1$ between $H$ and the exponent $h(2)$ for small scales.
In the case of fractional Gaussian noise, it was shown that
$H=h(2)$ \cite{Movahed06}. Hence we can use the value of $H$
calculated from $h(2)$ to detect the nature of long-range
correlation (LRC) in time series under the assumption of
fractional Gaussian noise or fractional Brownian motion.

\section{Complex networks: horizontal visibility graph for time series}

 \ \ \ \  A graph (or network) is a collection of nodes, which
denote the elements of a system, and links or edges, which
identify the relations or interactions among these elements.

Inspired by the concept of visibility \cite{Berg08}, Lacasa {\it
et al.} \cite{Lacasa08} proposed a simple computational method to
convert a time series into a network, known as visibility graph
(VG). It has been shown that time series structures are inherited
in the associated graphs, such as periodic, random, and fractal
series map into regular, random, and scale-free networks
respectively \cite{Lacasa08}. In addition, the degree
distributions of the VGs constructed from random series by a
uniform distribution in $[0, 1]$ have exponential tails $P(k)\sim
exp(-\lambda k)$ \cite{Lacasa08}. For this case, the greater the
value of $\lambda$ is, the faster attenuation of $P(k)$ is along
with $k$. In contrast, the degree distributions of the VGs
converted from fractional Brownian motions have power-law tails
$P(k)\sim k^{-\alpha}$ \cite{Lacasa09}.

Then a simplification of VG algorithm was proposed to map a time
series into a horizontal visibility graph (HVG) \cite{Luque09}. A
HVG is obtained from the mapping of a time series into a network
according to the following horizontal visibility criterion: Given
a time series $\{x_1, x_2, ... ,x_N\}$, two arbitrary data points
$x_{i}$ and $x_{j}$ in the time series have horizontal visibility,
and consequently become two connected nodes in the associated
graph, if any other data point $x_{n}$ such that $i<n<j$ fulfils
\begin{equation}
x_{i}, x_{j} > x_{n}.
\end{equation}
Thus a connected, unweighted network can be constructed from a
time series and is called its {\it horizontal visibility graph}
(HVG). In fact, the HVG of a given time series is always a
subgraph of its associated VG. Luque {\it et al.} \cite{Luque09}
have shown that the degree distribution of a HVG constructed from
any random series has an exponential form $P(k) =
(3/4)exp(-k\ln(3/2))$. Then Lacasa {\it et al.} \cite{Lacasa10}
used the horizontal visibility algorithm to characterize and
distinguish between correlated, uncorrelated and chaotic
processes. They showed that in every case the series maps into a
graph with exponential degree distribution $P(k)\sim exp(-\lambda
k)$, where the value of $\lambda$ characterizes the specific
process \cite{Lacasa10}. In this paper, we adopt the horizontal
visibility algorithm to convert the time series of energy,
pressure and volume into HVGs.

\section{Fractal analysis of horizontal visibility graph (complex networks)}

\ \ \ \  The random sequential box-covering algorithm proposed by
Kim {\it et al.} \cite{Kim2007}
  is an efficient algorithm for fractal analysis of complex networks.
Recently, our group used this algorithm to study the fractal
properties of a family of fractal networks and recurrence networks
constructed from fractional Brownian motions \cite{Liu14, Li14}.

For a given HVG or network, we let $N_{B}(r)$ be the minimum
number of boxes with size $r$ which are needed to cover the entire
network. If the power-law relation $N_{B}(r) \sim r^{-d_{B}}$
holds for some scaling-range of $r$ and constant $d_B$, we say
that the fractality exists in this network, and $d_B$ is called
its fractal dimension. So, in practice, we often obtain the
fractal dimension $d_{B}$ by fitting the linear relationship
between $N_{B}(r)$ and $r$ in a log-log plot.

Before using the random sequential box-covering algorithm to
calculate the fractal dimension of a network, we need to use
Floyd's algorithm \cite{Floyd62} or Dijkstra's algorithm of MatLab
toolbox to calculate the shortest-path distance matrix $D$ of this
network according to its adjacency matrix $A$. The random
sequential box-covering algorithm \cite{Kim2007} can be summarized
as follows.

\vspace{0.3cm} (I) Ensure that all nodes in the network are not
covered, and no node has been selected as the center of a box.

(II) Denote the size of the network as $N$. We set $t = 1,2,
\ldots ,T$; here we take $T=1000$. Then we rearrange the nodes
into $T=1000$ different random orders. That is to ensure that the
nodes of a network are randomly chosen as center nodes.

(III) Set the radius $r$ of boxes which will be used to cover the
nodes in the range $[1,d]$, where $d$ is the diameter of the
network (i.e. the longest distance between nodes in the network).

(IV) Treat the nodes of the $t$th kind of random orders that we
have obtained in (II) as the center of a box successively, then
search all the other nodes. If a node has a distance within $r$ to
the center node  and has not been covered yet, then cover it.

(V) If no more new nodes can be covered by this box, then we
abandon this box.

(VI) Repeat (IV) - (VI) until all the nodes are covered by the
corresponding boxes. We denote the number of boxes in this box
covering as $N(t,r)$.

(VII) Repeat steps (III) and (VI) for all the random orders to
find a box covering with minimal number of boxes $N(t,r)$. Then
denote this minimal number of boxes as $N_B(r/d)$.

(VIII)  For different $r$, repeat (III)-(VII). Then we use the
linear regression of $-\ln (N_B(r/d))$ vs $\ln(r/d)$ to estimate
the fractal dimension of the HVG or network.

\vspace{0.3cm}

\section{Results and discussion}

\ \ \ \  We use a list of 29 proteins (without ligands, RNA, or
DNA, greater than 100 amino acids) as in \cite{Rueda07} to
simulate their dynamics (see Table 1) by NAMD.

  Using NAMD with energy step size 200$fs$, protein structures
from PDB were used as starting points for 10-$ns$ production
trajectories, performed at constant pressure $(1 atm)$ and
temperature $(310K)$ using standard coupling schemes (the same in
all cases). CHARMM force fields were used in production runs.
Then, we ontained time series of bonded potential, angle
potential, dihedral potential, improper potential, electrostatic
potential, Van der Waals potential, kinetic energy, total energy,
potential energy, pressure, and volume. For example, we plotted
the 11 time series of protein 1A3H in Figures 1 to 3.

Then we performed MF-DFA on these time series. As an example, we
show how to estimate the exponents $h(q)$ for the bonded potential
time series of protein 1A3H in Figure 4. The numerical results
showed that the best fit occurs in the range $s \in [3, 22]$. So
we used this scaling-range to calculate the exponents $h(q)$. We
plotted the $h(q)$ curves for the energy, pressure and volume time
series of protein 1A3H in Figure 5 as examples. From the shape of
the $h(q)$ curves of these time series for all proteins, we found
that all time series considered here are multifractal. Therefore
these time series are complicated and cannot be characterized by a
single fractal dimension. In fact, they depict heterogeneous
phenomena and different regions of each time series have different
fractal properties \cite{Stanley1988}. In particular, the
numerical values of $h(2)$ (which is related to the Hurst exponent
$H$) are given in Table 2. The average values and standard
deviations of $h(2)$ for all 11 parameters of 29 proteins are also
listed in the bottom two rows of Table 2.

 For energy parameters, the energy values are in $kcal/mol$, bonded potential
$\sim 10^{3}$, angle potential $\sim 10^{3}$, dihedral potential
$\sim 10^{3}$, improper potential $\sim 10^{2}$, electrostatic
potential $\sim 10^{5}$, Van der Waals potential $\sim 10^{4}$,
kinetic energy $\sim 10^{4}$, total energy $\sim 10^{5}$, and
potential energy $\sim 10^{5}$. From Table 2, we can see that the
series of total energy and potential energy are non-stationary
($h(2)>1.0$) and anti-persistent ($0<H=h(2)-1<0.5$). The other
time series are stationary and persistent apart from that of
pressure (with $H\approx 0.5$ indicating the absence of long-range
correlation). Among bonded potential, angle potential, dihedral
potential, and improper potential, we can see that the LRC in the
series of angle potential is the strongest  and that in those of
improper potential is the weakest, and the LRC in those of
dihedral potential is stronger than that in those of bonded
potential. Among the three largest energy parameters, namely the
electrostatic potential, Vdw potential and kinetic energy, we find
that the LRC in the series of electrostatic potential is the
strongest and that in the series of Vdw potential is the weakest.

\textbf{Remark 1:} The NMR structures of many proteins in PDB were
obtained from different models. For example, the structures of
Protein 1WUZ in PDB were obtained from 30 NMR models. In this
case, we always use the structure obtained from the 1st NMR model
in the present study. We also tested the effect of different NMR
models on our results. We simulated the 1st model, 10th model,
20th model, and 30th model of protein 1WUZ. Our numerical results
of MF-DFA showed that the $h(q)$ curves for the four models are
very close. And it should be required in PDB that the structures
of any protein obtained from different NMR models should be
similar to each other. Therefore, it is unlikely that different
NMR models of proteins would affect our results.

Then, these time series were converted into HVGs. We denote $k$
the degree of nodes, $P(k)$ the probability of degree $k$. We
obtained the plots for the relations $k\sim P(k)$, $\ln(k)\sim
\ln(P(k))$, and $k\sim \ln(P(k))$, and found that only the linear
relationship in the $k\sim \ln(P(k))$ plots is reasonable.  We
gave the degree distributions $P(k)$ of the HVGs for the angle
potential and pressure time series of protein 1A3H in Figure 6 as
examples. So the HVGs of these parameters are exponential
networks. We gave the numerical values of the exponent $\lambda$
in Table 3. The average values and standard deviations of
$\lambda$ for all 11 parameters of the 29 proteins are also given
in Table 3. Among the three largest energy parameters, namely the
electrostatic potential, Vdw potential and kinetic energy, the
attenuation of the degree distribution for HVGs of electrostatic
potential is the fastest. The degree distribution for HVGs of
total energy has the fastest attenuation among all energy
parameters. The degree distribution for HVGs of potential energy
has the faster attenuation comparing to its components.

We also studied the fractal properties of the converted HVGs using
the random sequential box-covering algorithm described in Section
5. We found that there is a scaling-range of $r$ in which the
power-law relation $N_{B}(r) \sim r^{-d_{B}}$ holds for all
converted HVGs, hence the fractality exists in these networks. We
showed how to estimate the fractal dimension $d_B$ of HVGs
constructed from the electrostatic potential and van der Waals
potential time series of protein 1A3H in Figure 7 as examples. The
fractal dimension $d_{B}$ is the slope of linear regression
between $-\ln (N_{B}(r/d))$ and $\ln(r/d)$ for each case. In Table
4, we gave the fractal dimensions $d_{B}$ for all time series we
considered here. The bottom two rows of Table 4 are the average
values and standard deviations of $d_{B}$ for all 11 parameters of
29 proteins, respectively.

 From Tables 2, 3 and 4, for each energy or pressure or
volume parameter, one can find that the values of $h(2)$ of MF-DFA
on the time series, exponent $\lambda$ of exponential degree
distribution and fractal dimension $d_B$ of their converted HVGs
do not change much for different proteins (indicating some
universality). So these methods in this paper cannot be used for
the prediction of protein structures and functions.

For 29 time series (29 proteins) of each kind of energy or
pressure or volume, we calculated the average values of $h(2)$
(from MF-DFA on time series) and $d_{B}$ the converted HVGs. Hence
we obtained 11 values of $\langle h(2) \rangle$ and 11 values of
$\langle d_{B} \rangle$ respectively. It is surprising that there
is a nice linear relationship between $\langle h(2) \rangle$ and
$\langle d_{B} \rangle$ as shown in Figure 8. The relationship can
be well fitted by the linear formula:
$$\langle d_{B} \rangle = -1.5210\times\langle h(2) \rangle +3.1176.$$

\section{Conclusions}

\ \ \ \  Using the NAMD software, we derived the time series of
bonded potential, angle potential, dihedral potential, improper
potential, electrostatic potential, Van der Waals potential,
kinetic energy, total energy, potential energy, pressure, and
volume for each protein. The shape of the $h(q)$ curves from
MF-DFA indicates that these time series are multifractal.

 In particular, the numerical values of $h(2)$ of MF-DFA (which
is related to the Hurst exponent $H$) show that the series of
total energy and potential energy are non-stationary and
anti-persistent; the other time series are stationary and
persistent apart from the series of pressure (with $H\approx 0.5$
indicating the absence of long-range correlation).

The degree distributions of their converted HVGs show that these
networks are exponential. Among the three largest energy
parameters, namely the electrostatic potential, Vdw potential and
kinetic energy, the attenuation of the degree distribution of
electrostatic potential is the fastest. The degree distribution
for HVGs of total energy has the fastest attenuation among all
energy parameters. The degree distribution for HVGs of potential
energy has the faster attenuation comparing to its components.

We found that there is a scaling-range of $r$ in which the
power-law relation $N_{B}(r) \sim r^{-d_{B}}$ holds for all
converted HVGs, hence fractality exists in these networks.

For each energy, pressure or volume parameter, it is found that
the values of $h(2)$ of MF-DFA on the time series, exponent
$\lambda$ of exponential degree distribution and fractal dimension
$d_B$ of their converted HVGs do not change much for different
proteins (indicating some universality). So these methods cannot
be used for the prediction of protein structures and functions.

After taking average over all proteins, we surprisingly found that
there is a linear relationship between $\langle h(2) \rangle$
(from MF-DFA on time series) and $\langle d_{B} \rangle$ of
converted HVGs for different energy, pressure and volume.

\section*{ACKNOWLEDGEMENTS}

\ \ \ \ This project was supported by the Natural Science
Foundation of China (Grant No. 11371016); the Chinese Program for
Changjiang Scholars and Innovative Research Team in University
(PCSIRT) (Grant No. IRT1179); the Research Foundation of Education
Commission of Hunan Province of China (Grant No. 11A122); the
Lotus Scholars Program of Hunan province of China;  postdoctoral
research fund of Xiangtan University of China.

\newpage
\begin{table}
\caption{Structures representative of protein dynamics}
\centerline{\footnotesize
\begin{tabular}{|c|c|c|c|c|}
\hline \multicolumn{1}{|c}{PDB ID code} &\multicolumn{1}{|c}{The
number of amino acids} &\multicolumn{1}{|c}{Helical $(\%)$}
&\multicolumn{1}{|c}{Beta sheet $(\%)$} &\multicolumn{1}{|c|}{Exp
structure}\\\hline
   1A3H &  300 &  38 &  19 & X-ray\\
   1BFG &  126 &  8 &  35 & X-ray\\
   1BSN &  138 &  26 &  37 & NMR\\
   2DN8 &  100 &  0 &  37 & NMR\\
   2DO7 &  101 &  43 &  9 & NMR\\
   1F39 &  101 &  3 &  39 & X-ray\\
   1FE6 &  108 &  90 &  0 & X-ray\\
   1FPR &  294 &  30 &  19 & X-ray\\
   1H6T &  291 &  17 &  28 & X-ray\\
   1I4S &  294 &  68 &  1 & X-ray\\
   1IQQ &  200 &  29 &  20 & X-ray\\
   1JLI &  112 &  58 &  0 & NMR\\
   1K40 &  126 &  86 &  0 & X-ray\\
   1KS9 &  291 &  47 &  19 & X-ray\\
   1KTE &  105 &  50 &  19 & X-ray\\
   1KUU &  202 &  20 &  38 & X-ray\\
   1KXA &  159 &  5 &  47 & X-ray\\
   1MIX &  206 &  39 &  18 & X-ray\\
   1MJY &  350 &  20 &  33 & X-ray\\
   1N12 &  298 &  4 &  55 & X-ray\\
   1OO9 &  294 &  25 &  17 & NMR\\
   1OOI &  124 &  66 &  1 & X-ray\\
   1PHN &  334 &  76 &  0 & X-ray\\
   1RAL &  308 &  36 &  15 & X-ray\\
   1SUR &  215 &  48 &  14 & X-ray\\
   1WUZ &  103 &  12 &  34 & NMR\\
   1WWB &  103 &  2 &  49 & X-ray\\
   1X0M &  403 &  47 &  15 & X-ray\\
   1XGO &  295 &  30 &  22 & X-ray\\\hline
\end{tabular}}
\end{table}

\begin{table}
\caption{The exponent $h(2)$ of the time series for 29 proteins.
(Ave($h(2)$) and Std($h(2)$) denote the average and the standard
deviations of $h(2)$ for all 11 parameters of 29 proteins,
respectively.)}
 \centerline{\footnotesize
\begin{tabular}{|c|c|c|c|c|c|c|c|c|c|c|c|}
\hline \multicolumn{1}{|c}{PDB ID} &\multicolumn{1}{|c}{Bond-p}
&\multicolumn{1}{|c}{ang-p} &\multicolumn{1}{|c}{dih-p}
&\multicolumn{1}{|c}{imp-p} &\multicolumn{1}{|c}{ele-p}
&\multicolumn{1}{|c}{Vdw-p} &\multicolumn{1}{|c}{Kin-e}
&\multicolumn{1}{|c}{Tot-e} &\multicolumn{1}{|c}{Pot-e}
&\multicolumn{1}{|c}{pre} &\multicolumn{1}{|c|}{vol}\\\hline
   1A3H & 0.6266 & 0.7543 & 0.6880 & 0.5951 & 0.9052 & 0.6054 & 0.8340 & 1.2845 & 1.0454 & 0.4683 & 0.8988\\
   1BFG & 0.6269 & 0.7531 & 0.7201 & 0.5790 & 0.9401 & 0.6087 & 0.8431 & 1.2971 & 1.0685 & 0.4891 & 0.8646\\
   1BSN & 0.6227 & 0.7597 & 0.7317 & 0.5792 & 0.9283 & 0.5989 & 0.8350 & 1.3072 & 1.0590 & 0.4952 & 0.8721\\
   2DN8 & 0.6260 & 0.7589 & 0.7170 & 0.5848 & 0.9229 & 0.6227 & 0.8186 & 1.2910 & 1.0518 & 0.4960 & 0.8768\\
   2DO7 & 0.6254 & 0.7615 & 0.7028 & 0.5823 & 0.9344 & 0.6150 & 0.8332 & 1.3074 & 1.0630 & 0.5117 & 0.8934\\
   1F39 & 0.6334 & 0.7481 & 0.7116 & 0.5945 & 0.9267 & 0.6087 & 0.8316 & 1.3219 & 1.0593 & 0.5175 & 0.8887\\
   1FE6 & 0.6265 & 0.7512 & 0.7289 & 0.5821 & 0.9383 & 0.6009 & 0.8388 & 1.3158 & 1.0662 & 0.5172 & 0.8729\\
   1FPR & 0.6262 & 0.7562 & 0.7283 & 0.5616 & 0.9021 & 0.5948 & 0.8256 & 1.3071 & 1.0423 & 0.4762 & 0.9019\\
   1H6T & 0.6263 & 0.7551 & 0.7101 & 0.5928 & 0.9300 & 0.6282 & 0.8200 & 1.3063 & 1.0715 & 0.5062 & 0.8668\\
   1I4S & 0.6270 & 0.7237 & 0.7135 & 0.5753 & 0.9367 & 0.6002 & 0.8127 & 1.2979 & 1.0688 & 0.4874 & 0.8956\\
   1IQQ & 0.6265 & 0.7334 & 0.7353 & 0.5893 & 0.9225 & 0.5953 & 0.8383 & 1.3070 & 1.0701 & 0.4848 & 0.8762\\
   1JLI & 0.6276 & 0.7509 & 0.7070 & 0.5673 & 0.9296 & 0.5996 & 0.8438 & 1.2918 & 1.0450 & 0.4924 & 0.8657\\
   1K40 & 0.6253 & 0.7614 & 0.7295 & 0.5790 & 0.9278 & 0.6141 & 0.8236 & 1.2869 & 1.0648 & 0.5143 & 0.8899\\
   1KS9 & 0.6232 & 0.7621 & 0.7154 & 0.5844 & 0.9184 & 0.5912 & 0.8471 & 1.3158 & 1.0575 & 0.4762 & 0.8790\\
   1KTE & 0.6267 & 0.7597 & 0.7236 & 0.5877 & 0.9230 & 0.5937 & 0.8328 & 1.3090 & 1.0584 & 0.4725 & 0.8860\\
   1KUU & 0.6262 & 0.7599 & 0.7169 & 0.5761 & 0.9125 & 0.6029 & 0.8372 & 1.3036 & 1.0638 & 0.4824 & 0.8720\\
   1KXA & 0.6263 & 0.7684 & 0.7087 & 0.5835 & 0.9251 & 0.6196 & 0.8213 & 1.3132 & 1.0466 & 0.5200 & 0.8979\\
   1MIX & 0.6401 & 0.7643 & 0.7317 & 0.5921 & 0.9418 & 0.6290 & 0.8328 & 1.3234 & 1.0571 & 0.5342 & 0.8744\\
   1MJY & 0.6269 & 0.7665 & 0.6834 & 0.5711 & 0.9290 & 0.5903 & 0.8309 & 1.3001 & 1.0481 & 0.5120 & 0.8732\\
   1N12 & 0.6202 & 0.7530 & 0.7149 & 0.5872 & 0.9208 & 0.5965 & 0.8118 & 1.2997 & 1.0652 & 0.4719 & 0.9032\\
   1OO9 & 0.6250 & 0.7639 & 0.7081 & 0.5768 & 0.9201 & 0.5907 & 0.8310 & 1.3138 & 1.0761 & 0.4779 & 0.8849\\
   1OOI & 0.6277 & 0.7565 & 0.7014 & 0.5894 & 0.9148 & 0.6012 & 0.8324 & 1.2887 & 1.0643 & 0.4757 & 0.8916\\
   1PHN & 0.6317 & 0.7603 & 0.7061 & 0.5759 & 0.9308 & 0.6022 & 0.8339 & 1.3158 & 1.0470 & 0.4898 & 0.8814\\
   1RAL & 0.6277 & 0.7580 & 0.7222 & 0.5731 & 0.9366 & 0.5940 & 0.8344 & 1.2934 & 1.0650 & 0.4822 & 0.8810\\
   1SUR & 0.6219 & 0.7569 & 0.7242 & 0.5680 & 0.9318 & 0.6038 & 0.8322 & 1.3156 & 1.0577 & 0.4903 & 0.8772\\
   1WUZ & 0.6340 & 0.7606 & 0.7163 & 0.5770 & 0.9274 & 0.6067 & 0.8294 & 1.3236 & 1.0575 & 0.5172 & 0.8824\\
   1WWB & 0.6555 & 0.7572 & 0.7164 & 0.5916 & 0.9274 & 0.6213 & 0.8339 & 1.3104 & 1.0717 & 0.5005 & 0.8710\\
   1X0M & 0.6205 & 0.7513 & 0.7102 & 0.5885 & 0.9172 & 0.6072 & 0.8195 & 1.2950 & 1.0460 & 0.4925 & 0.8712\\
   1XGO & 0.6103 & 0.7550 & 0.7043 & 0.5779 & 0.9335 & 0.6051 & 0.8274 & 1.2969 & 1.0691 & 0.4836 & 0.8958\\\hline\hline
   Ave($h(2)$) & 0.6272 & 0.7559 & 0.7147 & 0.5815 & 0.9260 & 0.6051 & 0.8306 & 1.3049 & 1.0595 & 0.4943 & 0.8823\\
   Std($h(2)$) & 0.0074 & 0.0090 & 0.0123 & 0.0086 & 0.0096 & 0.0110 & 0.0086 & 0.0111 & 0.0095 & 0.0174 & 0.0114\\\hline
\end{tabular}}
\end{table}

\begin{table}
\caption{The exponent $\lambda$ of the horizontal visibility
graphs of the time series for 29 proteins. (Ave($\lambda$) and
Std($\lambda$) denote the average and the standard deviations of
$\lambda$ for all 11 parameters of 29 proteins, respectively.)}
\centerline{\footnotesize
\begin{tabular}{|c|c|c|c|c|c|c|c|c|c|c|c|}
\hline \multicolumn{1}{|c}{PDB ID} &\multicolumn{1}{|c}{Bond-p}
&\multicolumn{1}{|c}{ang-p} &\multicolumn{1}{|c}{dih-p}
&\multicolumn{1}{|c}{imp-p} &\multicolumn{1}{|c}{ele-p}
&\multicolumn{1}{|c}{Vdw-p} &\multicolumn{1}{|c}{Kin-e}
&\multicolumn{1}{|c}{Tot-e} &\multicolumn{1}{|c}{Pot-e}
&\multicolumn{1}{|c}{pre} &\multicolumn{1}{|c|}{vol}\\\hline
   1A3H & 0.4139 & 0.4398 & 0.4236 & 0.4129 & 0.4422 & 0.4163 & 0.4282 & 0.6238 & 0.4719 & 0.4068 & 0.5088\\
   1BFG & 0.4115 & 0.4386 & 0.4274 & 0.4092 & 0.4469 & 0.4172 & 0.4293 & 0.6341 & 0.4737 & 0.4095 & 0.5032\\
   1BSN & 0.4176 & 0.4333 & 0.4276 & 0.4149 & 0.4413 & 0.4166 & 0.4270 & 0.6273 & 0.4748 & 0.4098 & 0.5086\\
   2DN8 & 0.4120 & 0.4331 & 0.4264 & 0.4198 & 0.4442 & 0.4120 & 0.4309 & 0.6177 & 0.4749 & 0.4108 & 0.5038\\
   2DO7 & 0.4132 & 0.4365 & 0.4301 & 0.4145 & 0.4432 & 0.4162 & 0.4240 & 0.6131 & 0.4718 & 0.4070 & 0.5068\\
   1F39 & 0.4109 & 0.4390 & 0.4283 & 0.4113 & 0.4453 & 0.4178 & 0.4290 & 0.6243 & 0.4776 & 0.4101 & 0.5095\\
   1FE6 & 0.4068 & 0.4325 & 0.4167 & 0.4177 & 0.4442 & 0.4111 & 0.4256 & 0.6239 & 0.4776 & 0.4088 & 0.5033\\
   1FPR & 0.4188 & 0.4313 & 0.4298 & 0.4158 & 0.4387 & 0.4190 & 0.4306 & 0.6229 & 0.4752 & 0.3968 & 0.5015\\
   1H6T & 0.4188 & 0.4333 & 0.4208 & 0.4121 & 0.4428 & 0.4123 & 0.4230 & 0.6239 & 0.4749 & 0.4106 & 0.5066\\
   1I4S & 0.4102 & 0.4317 & 0.4238 & 0.4145 & 0.4403 & 0.4225 & 0.4247 & 0.6299 & 0.4745 & 0.4066 & 0.4853\\
   1IQQ & 0.4134 & 0.4315 & 0.4245 & 0.4176 & 0.4473 & 0.4193 & 0.4238 & 0.6247 & 0.4780 & 0.4060 & 0.5046\\
   1JLI & 0.4106 & 0.4152 & 0.4224 & 0.4157 & 0.4425 & 0.4191 & 0.4286 & 0.6330 & 0.4771 & 0.4074 & 0.4772\\
   1K40 & 0.4161 & 0.4329 & 0.4286 & 0.4158 & 0.4485 & 0.4142 & 0.4240 & 0.6213 & 0.4705 & 0.4036 & 0.5059\\
   1KS9 & 0.4128 & 0.4390 & 0.4208 & 0.4188 & 0.4427 & 0.4173 & 0.4231 & 0.6296 & 0.4717 & 0.4048 & 0.5048\\
   1KTE & 0.4188 & 0.4389 & 0.4268 & 0.4141 & 0.4486 & 0.4191 & 0.4243 & 0.6292 & 0.4763 & 0.4060 & 0.5073\\
   1KUU & 0.4132 & 0.4321 & 0.4292 & 0.4143 & 0.4428 & 0.4183 & 0.4266 & 0.6264 & 0.4792 & 0.4030 & 0.5035\\
   1KXA & 0.4111 & 0.4254 & 0.4285 & 0.4197 & 0.4352 & 0.4126 & 0.4240 & 0.6397 & 0.4733 & 0.4032 & 0.5033\\
   1MIX & 0.4163 & 0.4332 & 0.4266 & 0.4199 & 0.4490 & 0.4193 & 0.4265 & 0.6381 & 0.4794 & 0.4086 & 0.5041\\
   1MJY & 0.4176 & 0.4326 & 0.4254 & 0.4142 & 0.4430 & 0.4197 & 0.4231 & 0.6128 & 0.4793 & 0.4079 & 0.5028\\
   1N12 & 0.4174 & 0.4236 & 0.4206 & 0.4180 & 0.4477 & 0.4197 & 0.4260 & 0.6276 & 0.4729 & 0.4022 & 0.5081\\
   1OO9 & 0.4182 & 0.4371 & 0.4210 & 0.4094 & 0.4422 & 0.4170 & 0.4220 & 0.6297 & 0.4704 & 0.4058 & 0.5067\\
   1OOI & 0.4144 & 0.4315 & 0.4212 & 0.4161 & 0.4408 & 0.4162 & 0.4266 & 0.6212 & 0.4753 & 0.4085 & 0.5027\\
   1PHN & 0.4133 & 0.4255 & 0.4229 & 0.4132 & 0.4422 & 0.4186 & 0.4250 & 0.6328 & 0.4719 & 0.3952 & 0.5037\\
   1RAL & 0.4173 & 0.4303 & 0.4255 & 0.4166 & 0.4437 & 0.4103 & 0.4221 & 0.6308 & 0.4744 & 0.4003 & 0.4936\\
   1SUR & 0.4145 & 0.4350 & 0.4278 & 0.4164 & 0.4445 & 0.4189 & 0.4259 & 0.6439 & 0.4787 & 0.4076 & 0.4905\\
   1WUZ & 0.4152 & 0.4300 & 0.4265 & 0.4189 & 0.4423 & 0.4158 & 0.4217 & 0.6046 & 0.4747 & 0.4076 & 0.5063\\
   1WWB & 0.4103 & 0.4361 & 0.4231 & 0.4151 & 0.4460 & 0.4174 & 0.4237 & 0.6269 & 0.4799 & 0.4089 & 0.5065\\
   1X0M & 0.4155 & 0.4339 & 0.4246 & 0.4102 & 0.4426 & 0.4181 & 0.4268 & 0.6210 & 0.4778 & 0.4089 & 0.4953\\
   1XGO & 0.4169 & 0.4309 & 0.4215 & 0.4144 & 0.4447 & 0.4141 & 0.4227 & 0.6282 & 0.4712 & 0.4094 & 0.5020\\\hline\hline
   Ave($\lambda$) & 0.4144 & 0.4325 & 0.4249 & 0.4152 & 0.4436 & 0.4168 & 0.4255 & 0.6263 & 0.4751 & 0.4063 & 0.5023\\
   Std($\lambda$) & 0.0031 & 0.0052 & 0.0033 & 0.0029 & 0.0030 & 0.0029 & 0.0026 & 0.0081 & 0.0029 & 0.0038 & 0.0073\\\hline
\end{tabular}}
\end{table}

\begin{table}
\caption{The fractal dimension $d_{B}$ of the horizontal
visibility graph of the time series for 29 proteins. (Ave($d_{B}$)
and Std($d_{B}$) denote the average and the standard deviations of
$d_{B}$ for all 11 parameters of 29 proteins, respectively.)}
\centerline{\footnotesize
\begin{tabular}{|c|c|c|c|c|c|c|c|c|c|c|c|}
\hline \multicolumn{1}{|c}{PDB ID} &\multicolumn{1}{|c}{Bond-p}
&\multicolumn{1}{|c}{ang-p} &\multicolumn{1}{|c}{dih-p}
&\multicolumn{1}{|c}{imp-p} &\multicolumn{1}{|c}{ele-p}
&\multicolumn{1}{|c}{Vdw-p} &\multicolumn{1}{|c}{Kin-e}
&\multicolumn{1}{|c}{Tot-e} &\multicolumn{1}{|c}{Pot-e}
&\multicolumn{1}{|c}{pre} &\multicolumn{1}{|c|}{vol}\\\hline
   1A3H & 2.2420 & 1.9094 & 2.0924 & 2.1066 & 1.6176 & 2.2444 & 1.8359 & 1.1711 & 1.4636 & 2.3378 & 1.7238\\
   1BFG & 2.1676 & 1.9819 & 2.0500 & 2.2011 & 1.6010 & 2.2705 & 1.8846 & 1.1743 & 1.4867 & 2.2933 & 1.6836\\
   1BSN & 2.2406 & 1.9308 & 2.0763 & 2.1631 & 1.6173 & 2.2659 & 1.8608 & 1.1663 & 1.4840 & 2.3774 & 1.6850\\
   2DN8 & 2.1980 & 1.9635 & 2.1620 & 2.1677 & 1.6223 & 2.2684 & 1.8451 & 1.1764 & 1.4928 & 2.3181 & 1.8107\\
   2DO7 & 2.1975 & 1.9583 & 2.0344 & 2.1446 & 1.6304 & 2.3466 & 1.8794 & 1.1816 & 1.5028 & 2.3746 & 1.7386\\
   1F39 & 2.2468 & 1.9891 & 2.1464 & 2.1154 & 1.5909 & 2.2725 & 1.8349 & 1.1857 & 1.4706 & 2.2654 & 1.6819\\
   1FE6 & 2.2630 & 2.0082 & 2.0136 & 2.1567 & 1.6369 & 2.3287 & 1.8510 & 1.1717 & 1.4698 & 2.2609 & 1.7250\\
   1FPR & 2.2390 & 1.9955 & 2.1629 & 2.2214 & 1.6071 & 2.3019 & 1.8316 & 1.1665 & 1.4861 & 2.3601 & 1.7634\\
   1H6T & 2.3016 & 1.9043 & 2.0813 & 2.1474 & 1.6370 & 2.2001 & 1.8580 & 1.1864 & 1.4765 & 2.3218 & 1.7400\\
   1I4S & 2.2240 & 2.0261 & 2.0840 & 2.0849 & 1.6208 & 2.2055 & 1.8901 & 1.1820 & 1.5158 & 2.3245 & 1.7718\\
   1IQQ & 2.2468 & 1.9935 & 2.0428 & 2.2282 & 1.6163 & 2.2704 & 1.8316 & 1.1674 & 1.4841 & 2.4346 & 1.7014\\
   1JLI & 2.2892 & 2.0062 & 2.0662 & 2.1833 & 1.6201 & 2.3098 & 1.8878 & 1.1816 & 1.4788 & 2.2764 & 1.7530\\
   1K40 & 2.1872 & 1.9844 & 2.0558 & 2.1804 & 1.6435 & 2.2935 & 1.8767 & 1.1773 & 1.4626 & 2.4219 & 1.7508\\
   1KS9 & 2.2384 & 1.9884 & 2.0947 & 2.0910 & 1.6336 & 2.3056 & 1.8368 & 1.1753 & 1.4988 & 2.3386 & 1.7253\\
   1KTE & 2.1812 & 2.0072 & 2.0802 & 2.0754 & 1.6411 & 2.2450 & 1.8494 & 1.1789 & 1.4744 & 2.3078 & 1.7542\\
   1KUU & 2.1128 & 2.0287 & 2.1360 & 2.1860 & 1.6510 & 2.2234 & 1.8746 & 1.1909 & 1.4877 & 2.3736 & 1.7157\\
   1KXA & 2.1854 & 1.9436 & 2.0970 & 2.1582 & 1.6303 & 2.3596 & 1.8813 & 1.1700 & 1.5108 & 2.2604 & 1.7123\\
   1MIX & 2.2149 & 2.0408 & 2.1175 & 2.1299 & 1.6383 & 2.3023 & 1.8498 & 1.1522 & 1.4852 & 2.3154 & 1.7070\\
   1MJY & 2.1973 & 1.9703 & 2.0800 & 2.1877 & 1.6672 & 2.2188 & 1.9019 & 1.1565 & 1.4943 & 2.3145 & 1.7069\\
   1N12 & 2.2549 & 2.0646 & 2.0082 & 2.0736 & 1.6065 & 2.2561 & 1.8471 & 1.1778 & 1.4863 & 2.3750 & 1.7865\\
   1OO9 & 2.2266 & 2.0059 & 2.0746 & 2.0948 & 1.6369 & 2.3035 & 1.8906 & 1.1750 & 1.4822 & 2.3019 & 1.7576\\
   1OOI & 2.2811 & 1.9514 & 2.1226 & 2.1475 & 1.6275 & 2.2415 & 1.8729 & 1.1746 & 1.4662 & 2.2810 & 1.7851\\
   1PHN & 2.2882 & 1.9190 & 2.0631 & 2.1411 & 1.6152 & 2.3023 & 1.8914 & 1.1673 & 1.4838 & 2.4080 & 1.7280\\
   1RAL & 2.1851 & 2.0142 & 2.0490 & 2.0774 & 1.6226 & 2.2371 & 1.8833 & 1.1743 & 1.4981 & 2.3651 & 1.7658\\
   1SUR & 2.2318 & 2.0036 & 2.0301 & 2.0664 & 1.6251 & 2.2950 & 1.9093 & 1.1901 & 1.4917 & 2.2883 & 1.7264\\
   1WUZ & 2.2801 & 1.9021 & 2.0407 & 2.1896 & 1.6406 & 2.3845 & 1.8850 & 1.1693 & 1.5043 & 2.4449 & 1.6824\\
   1WWB & 2.2323 & 1.9986 & 2.0172 & 2.1306 & 1.6187 & 2.3751 & 1.8956 & 1.1743 & 1.4875 & 2.2518 & 1.7267\\
   1X0M & 2.2126 & 2.0268 & 2.0530 & 2.1421 & 1.6553 & 2.2801 & 1.8780 & 1.1676 & 1.4813 & 2.3462 & 1.7104\\
   1XGO & 2.1929 & 2.0190 & 2.0762 & 2.1694 & 1.6323 & 2.3051 & 1.8543 & 1.1876 & 1.4716 & 2.3140 & 1.7302\\\hline\hline
   Ave($d_{B}$) & 2.2262 & 1.9840 & 2.0761 & 2.1435 & 1.6277 & 2.2832 & 1.8695 & 1.1748 & 1.4855 & 2.3328 & 1.7327\\
   Std($d_{B}$) & 0.0424 & 0.0422 & 0.0419 & 0.0456 & 0.0164 & 0.0474 & 0.0233 & 0.0090 & 0.0134 & 0.0533 & 0.0331\\\hline
\end{tabular}}
\end{table}

\begin{figure}
\centerline{\epsfxsize=12cm \epsfbox{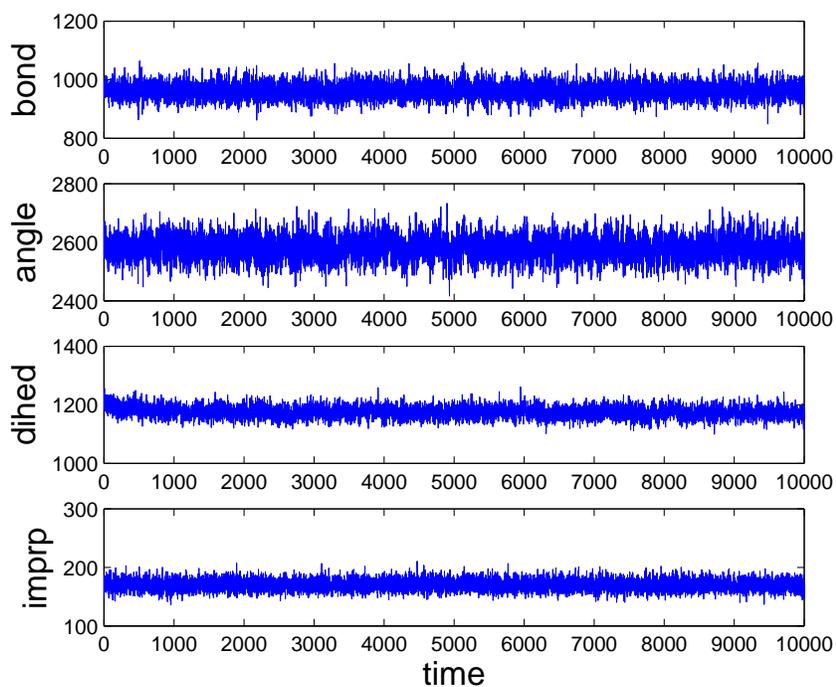}}
\caption{Bonded potential, angle potential, dihedral potential and
improper potential time series for protein 1A3H.} \label{g21}
\end{figure}

\begin{figure}
\centerline{\epsfxsize=12cm \epsfbox{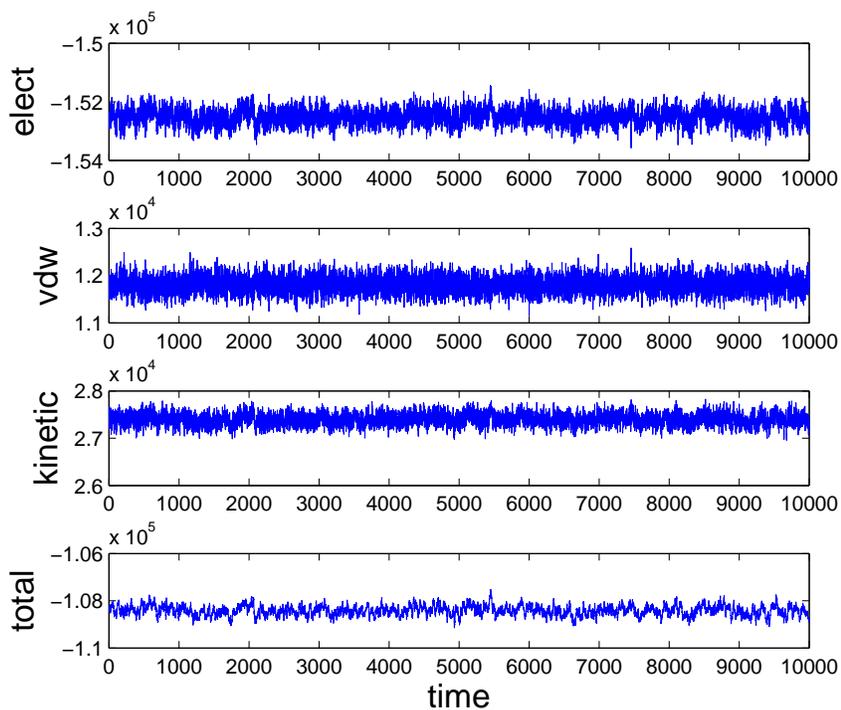}}
\caption{Electrostatic potential, Van der Waals potential, kinetic
energy and total energy time series for protein 1A3H.} \label{g22}
\end{figure}

\begin{figure}
\centerline{\epsfxsize=12cm \epsfbox{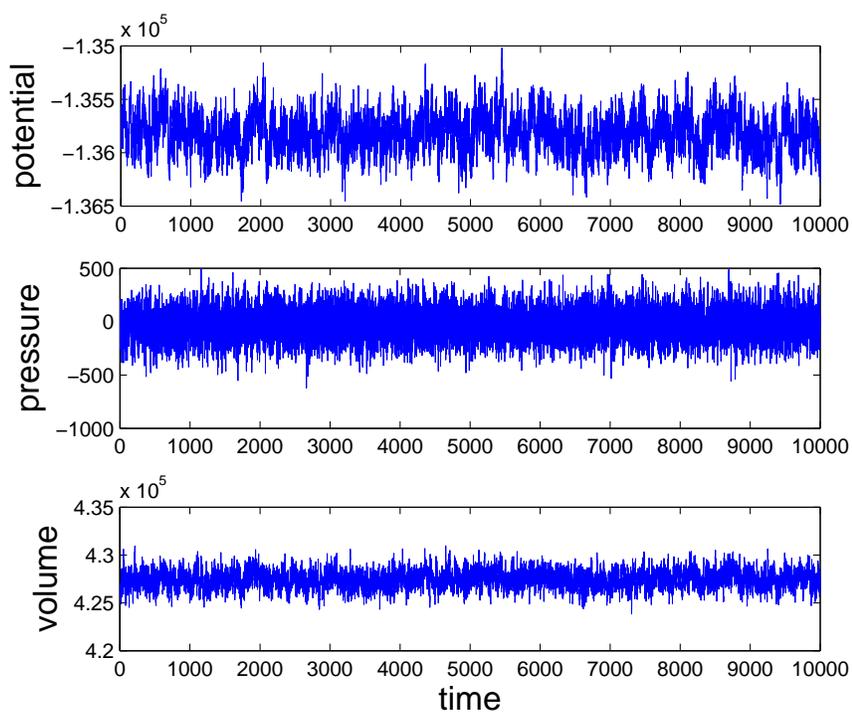}}
\caption{Potential energy, pressure, and volume time series for
protein 1A3H.} \label{g23}
\end{figure}

\begin{figure}
\centerline{\epsfxsize=12cm \epsfbox{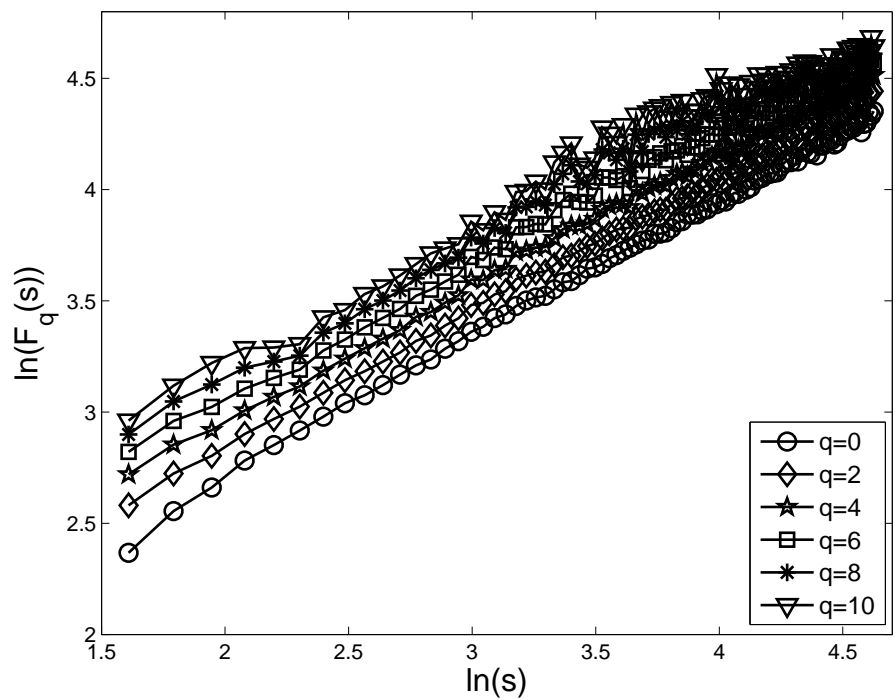}}
\caption{Linear regressions for calculating the exponents $h(q)$
for the bonded potential time series of protein 1A3H.} \label{g24}
\end{figure}

\begin{figure}
\centerline{\epsfxsize=12cm \epsfbox{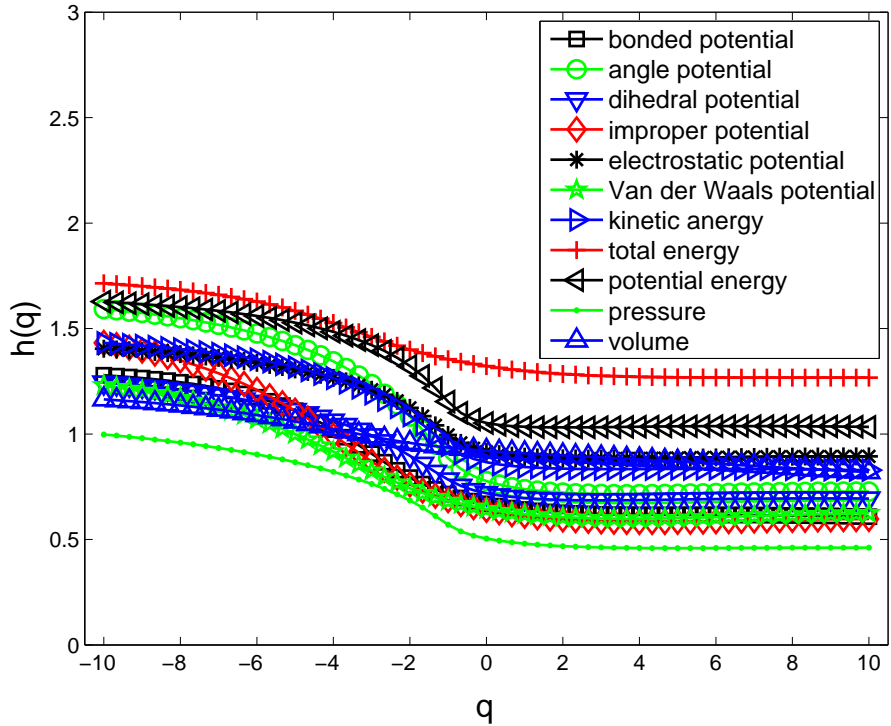}}
\caption{Multifractal $h(q)$ curves for the energy, pressure and
volume time series of protein 1A3H.} \label{g25}
\end{figure}

\begin{figure}
\centerline{\epsfxsize=12cm \epsfbox{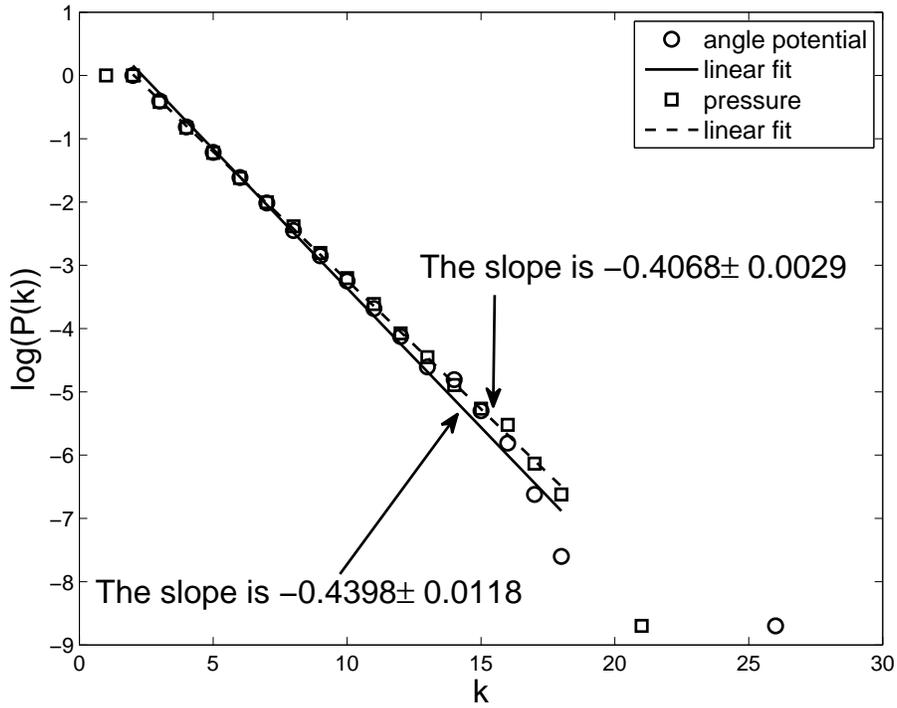}}
\caption{The degree distributions $P(k)$ of the horizontal
visibility graph for the angle potential and pressure time series
of protein 1A3H.} \label{g26}
\end{figure}

\begin{figure}
\centerline{\epsfxsize=12cm \epsfbox{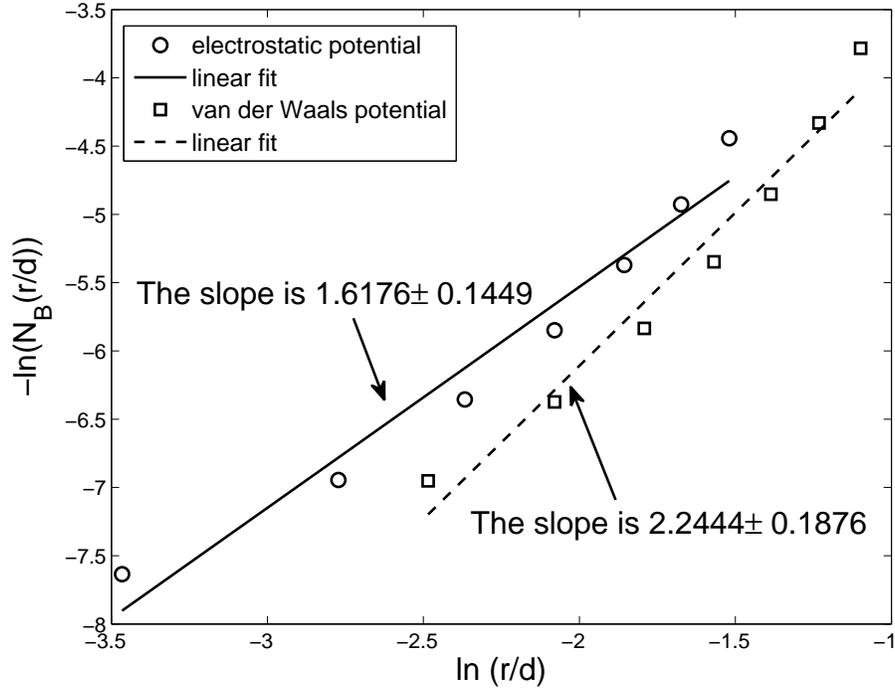}}
\caption{Linear regressions for calculating the fractal dimensions
$d_{B}$ of the horizontal visibility graphs for the electrostatic
potential and van der Waals potential time series of protein
1A3H.} \label{g27}
\end{figure}

\begin{figure}
\centerline{\epsfxsize=12cm \epsfbox{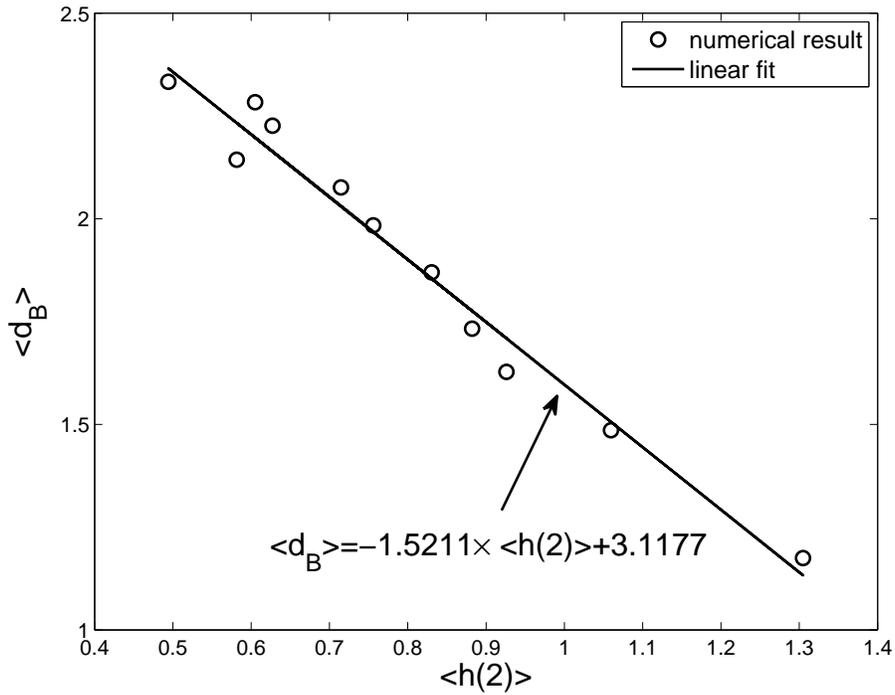}}
\caption{The relationship between $\langle h(2)\rangle$ (from
MF-DFA on time series) and $\langle d_{B} \rangle$ of the
constructed horizontal visibility graphs for different energy,
pressure and volume. The average is taken over 29 proteins.}
\label{g28}
\end{figure}

\end{document}